\documentclass[11pt]{article}
\usepackage{a4wide}
\usepackage[dvips]{graphicx}
\usepackage{times}
\usepackage{amsmath,amssymb}
\begin{document}
\title{%
Elusive multiquark spectroscopy 
\footnote{Invited talk at the 19th Few-Body Conference, Bonn, Germany, 2009, to appear in the Proceedings}
}%
\date{\small\today}
\author{%
Jean-Marc Richard\footnote{email: \textsf{j-m.richard@ipnl.in2p3.fr}} 
\\
{\small
Laboratoire de Physique Subatomique et Cosmologie,}\\[-2pt]
{\small IN2P3-CNRS, Universit\'e Joseph Fourier,  INPG,}\\[-2pt]
{\small 53, avenue des Martyrs, 38026 Grenoble, France,}\\[-2pt]
{\small and }\\[-2pt]
{\small Institut de Physique Nucl\'eaire de Lyon,
Universit\'e de Lyon, IN2P3-CNRS-UCBL,}\\[-2pt]
{\small 4, rue Enrico Fermi, 69622 Villeurbanne, France}\\[-2pt]
}
\maketitle

\begin{abstract}
A review is presented of past and recent attempts to build multiquark states within current models already describing ordinary mesons and baryons.
This includes:  coherence in the chromomagnetic interaction,  tetraquarks with two heavy quarks, Steiner-tree models of confinement, and  hadronic molecules, in particular in the hidden-charm sector. 
Some emphasis is put on  the difficulties encountered when extrapolating toward higher configurations the dynamics of confining forces, starting  from the simple case of a quark and an antiquark, or three quarks in a colour singlet.
\end{abstract}

%
%
\section{Introduction}
\label{RichardJM_intro}
In principle, especially in front of this audience of the International Few-Body Conference, the task of describing multiquark spectroscopy should be rather easy: a model is first tuned to fit the spectrum and properties of mesons and baryons, and then applied to multiquark configurations. The main effort would hence be on the subtleties of the few-quark problems, which might differ from other few-body systems by the confining character of the interaction.

However, the procedure is not that straightforward, because
 the quark dynamics governing mesons and baryons is not very well known, and, within any assumed model, the extrapolation towards multiquarks involves severe uncertainties.

This talk is organised as follows. In Sec.~\ref{se:history}, some milestones in the multiquark history are reminded. In Sec.~\ref{se:quarks}, binding mechanisms are discussed in the framework of quark dynamics, with emphasis on spin-independent forces.  In Sec.~\ref{se:Steiner}, the Steiner-tree model of confinement is applied to multiquarks. Section~\ref{se:out} is devoted to some conclusions.

Due to space limitations, the list of references will  necessarily be limited. Apologies are presented to the many colleagues whose interesting work is implicitly taken into account, but not explicitly cited.

\section{History}\label{se:history}
\subsection{Brief survey}
The search for exotic hadrons started even before the quark model, though the ``exotic character'' was not very precisely defined. 
For instance, it was observed that hyperons with baryon number $B=+1$ have a negative strangeness number $S$, once $S=+1$ is arbitrarily attributed to $K^+$. Some indications came, however,  on a baryon with $S=+1$, the so-called ``$Z$'' resonance. See, e.g., early issues of \emph{Review of particle properties}, cited in the last one \cite{Amsler:2008zz}. Eventually, the analysis of the $KN$ data did not confirm the existence of the $Z$ baryon.  Regularly, indications have been reported about ``dibaryon'' resonances, but, again, no firm confirmation was ever reached.  

Today, by ``exotic'', one means a state whose quantum numbers and main properties cannot be matched by a quark--antiquark $(q\bar{q})$ or  a three-quark $(qqq)$ configuration. 
In explicit quark models,  and in the data, the indications are often in a sector of \emph{crypto-exotic} states, with the same quantum numbers as for ordinary hadrons, but intriguing properties. For instance, a four-quark state with isospin $I=0$ and spin-parity $J^P=0^+$ competes with an orbital excitation of $(q\bar{q})$ where a spin triplet couples to an orbital momentum $L=1$, to form  $J^P=0^+$. The problem is whether there is evidence for supernumerary states. Otherwise, the dominant $(q\bar{q})$ structure is simply modified by higher Fock components.

The experimental situation had unfortunately ups and downs: baryonium, dibaryon, pentaquark or molecule candidates have been reported, and usually not confirmed. Some scepticism tends thus to prevail among non-experts. On the theory side, also, the situation is seen by outsiders as rather confuse, with promising scenarios for multiquark binding often not confirmed by more detailed studies.

Diverse mechanisms have, indeed, been proposed for building hadrons beyond $(q\bar{q})$ or $(qqq)$: chromomagnetism, chiral dynamics, favourable symmetry breaking, etc.,  but none of them has been fully successful and convincing, so far. There is also a recurrent problem attached to almost any idea about multiquark binding: once a state is proposed, with due caution, followers jump on the idea and list dozens of ``states", most of them lying well above their lowest dissociation threshold, and thus likely falling apart immediately and  not visible as peaks. 

For example, beautiful mass formulas have been written in the late 70's to describe hadrons made of clusters whose mass is shifted by colour-dependent hyperfine for\-ces.  These formulas were elaborated using sophisticated group-theoretical techniques. However, 
estimating hyperfine effects does not address the problem of resisting spontaneous fall-apart into two colour singlets. Hence any multiquark above the dissociation threshold should be considered with a lot of caution. Jaffe and Low \cite{Jaffe:1978bu} made this point more rigorous. Multiquark states with an artificial confinement constitutes a basis of ``primitives'' for describing the short-range hadron--hadron dynamics, and usually have little to do with candidates for new states in the spectrum.

\subsection{Exchange of particles}
\begin{figure}[!bhtc]
\begin{center}
\vskip -.4cm
\includegraphics[width=.24\columnwidth]{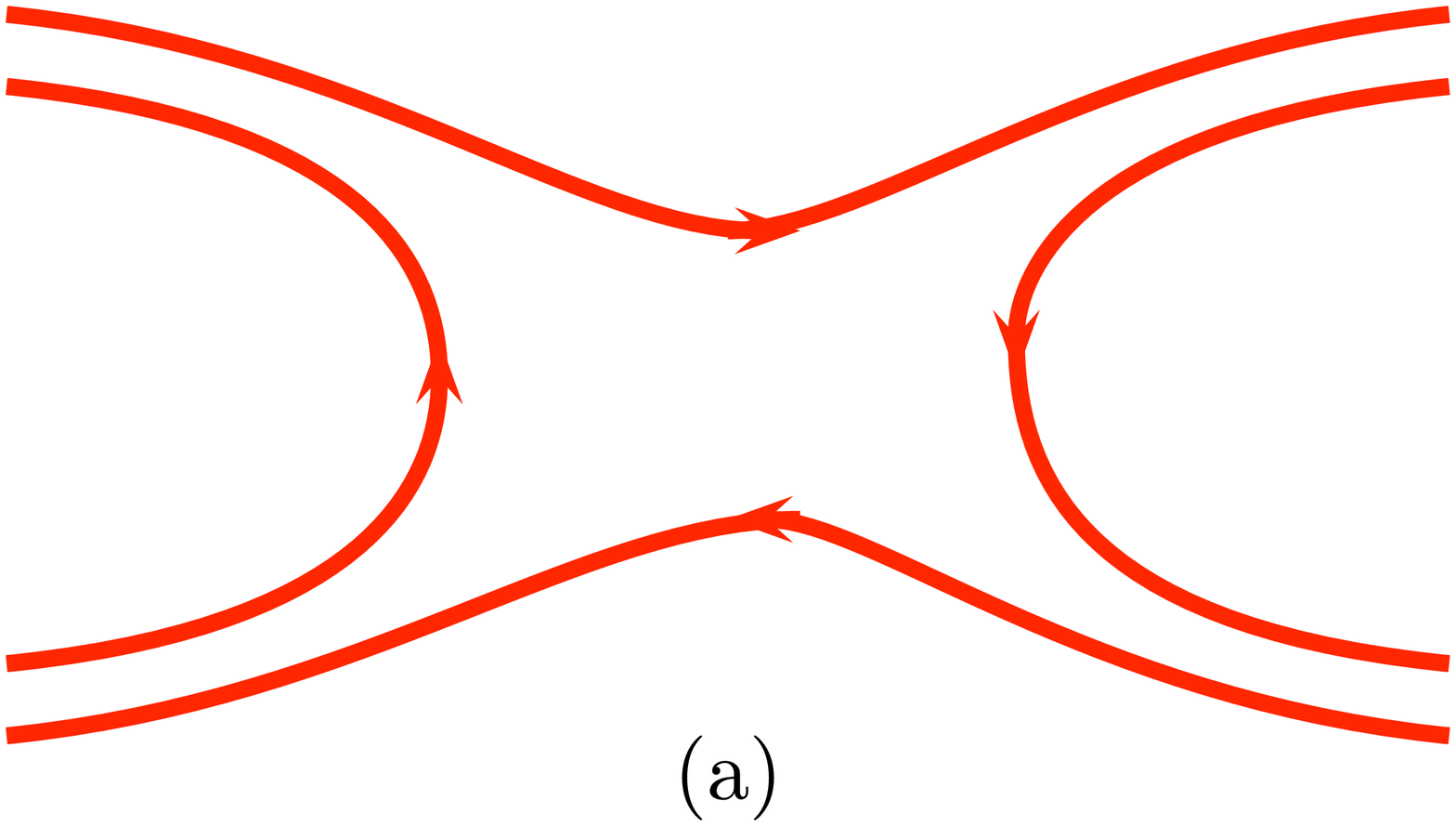}\qquad
\includegraphics[width=.24\columnwidth]{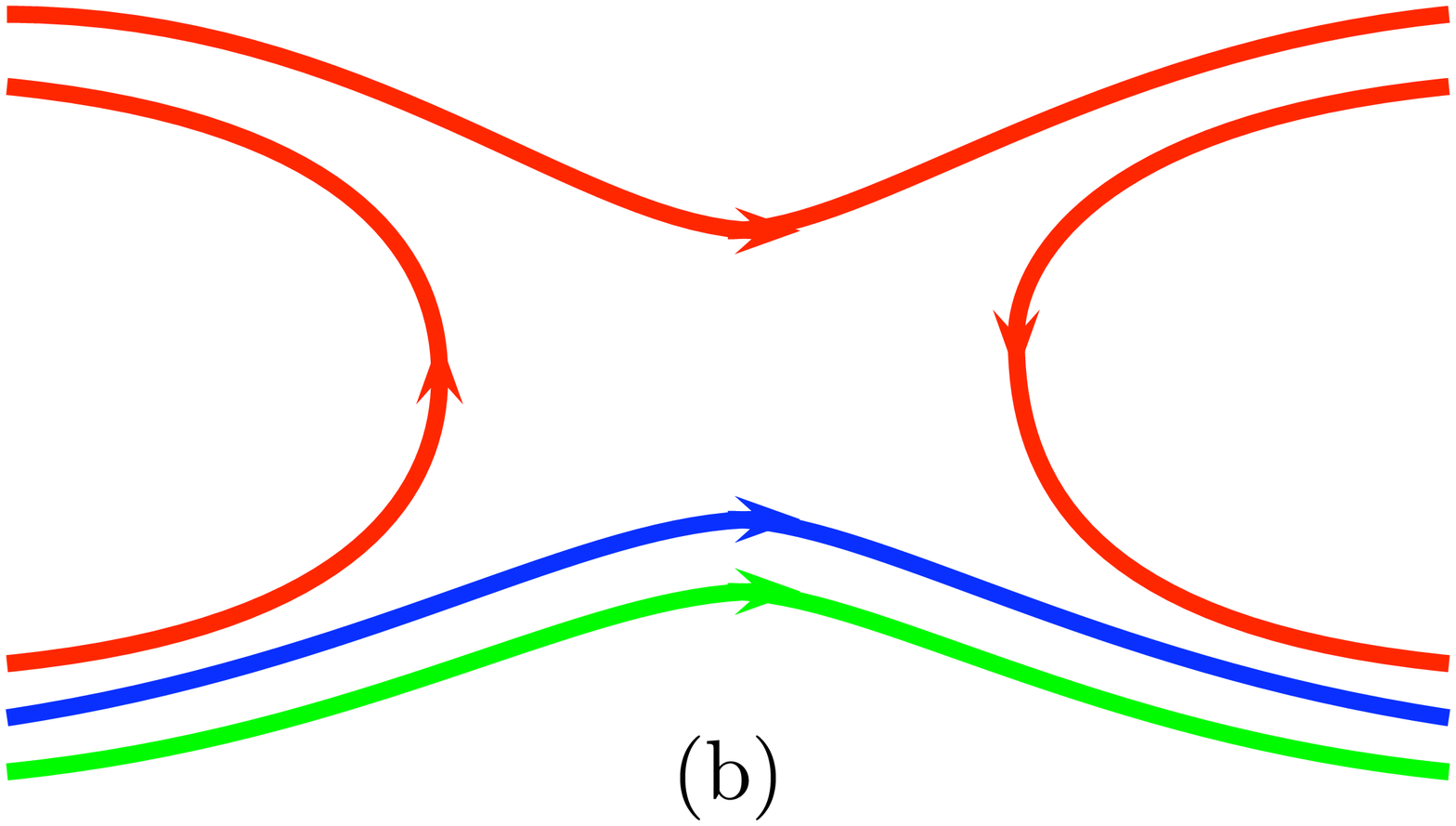}\qquad
\includegraphics[width=.24\columnwidth]{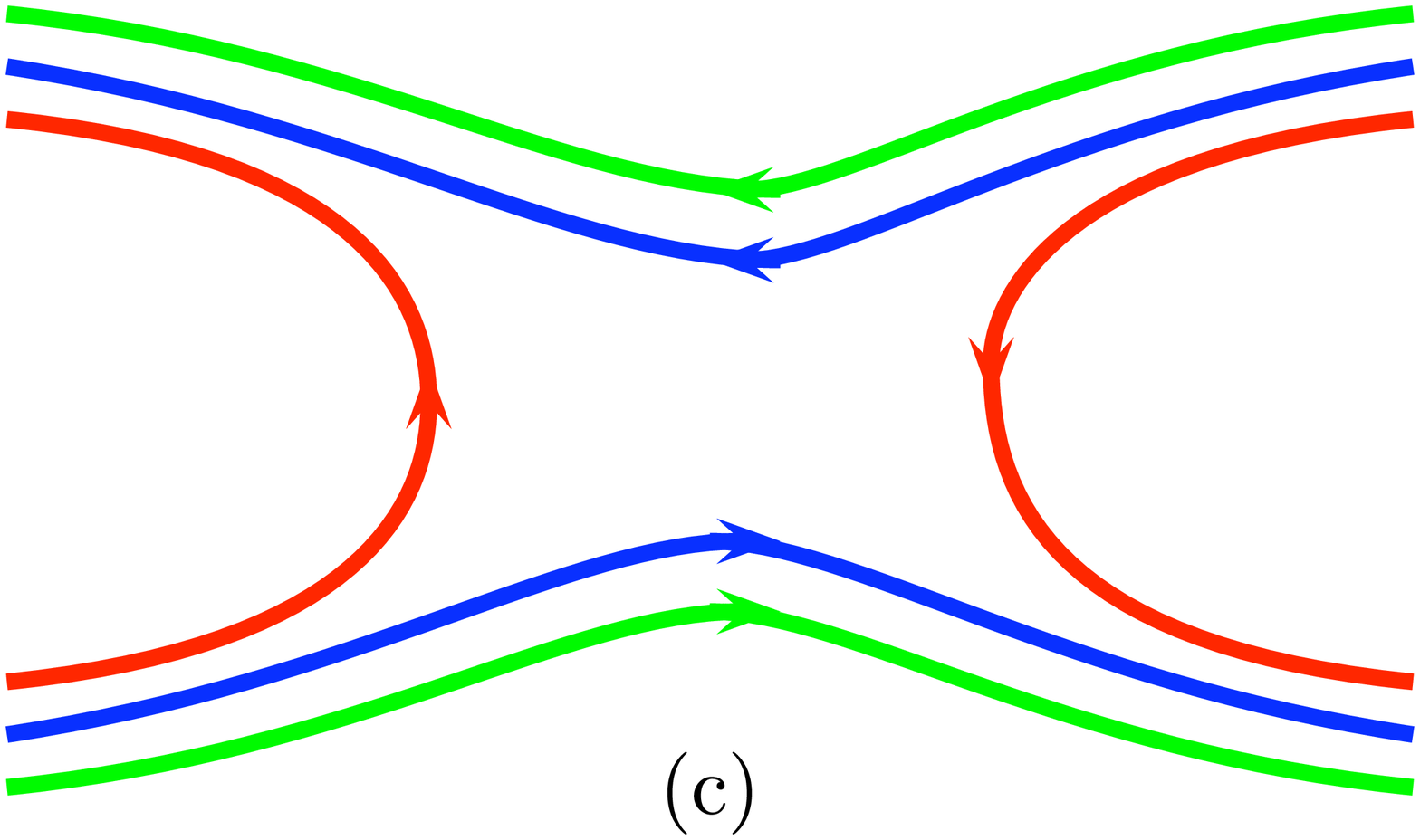}\\[5pt]
\includegraphics[width=.24\columnwidth]{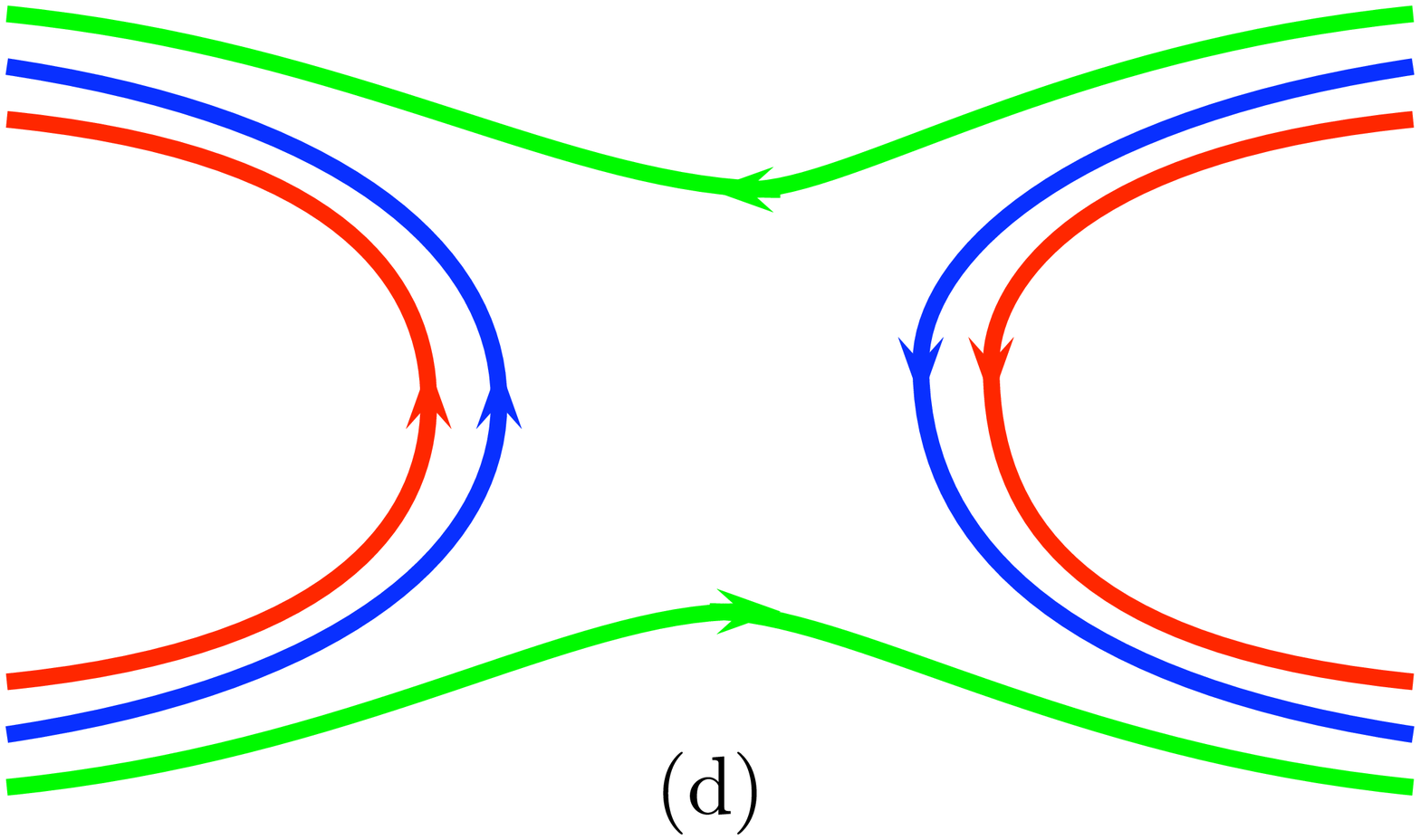}\qquad
\includegraphics[width=.24\columnwidth]{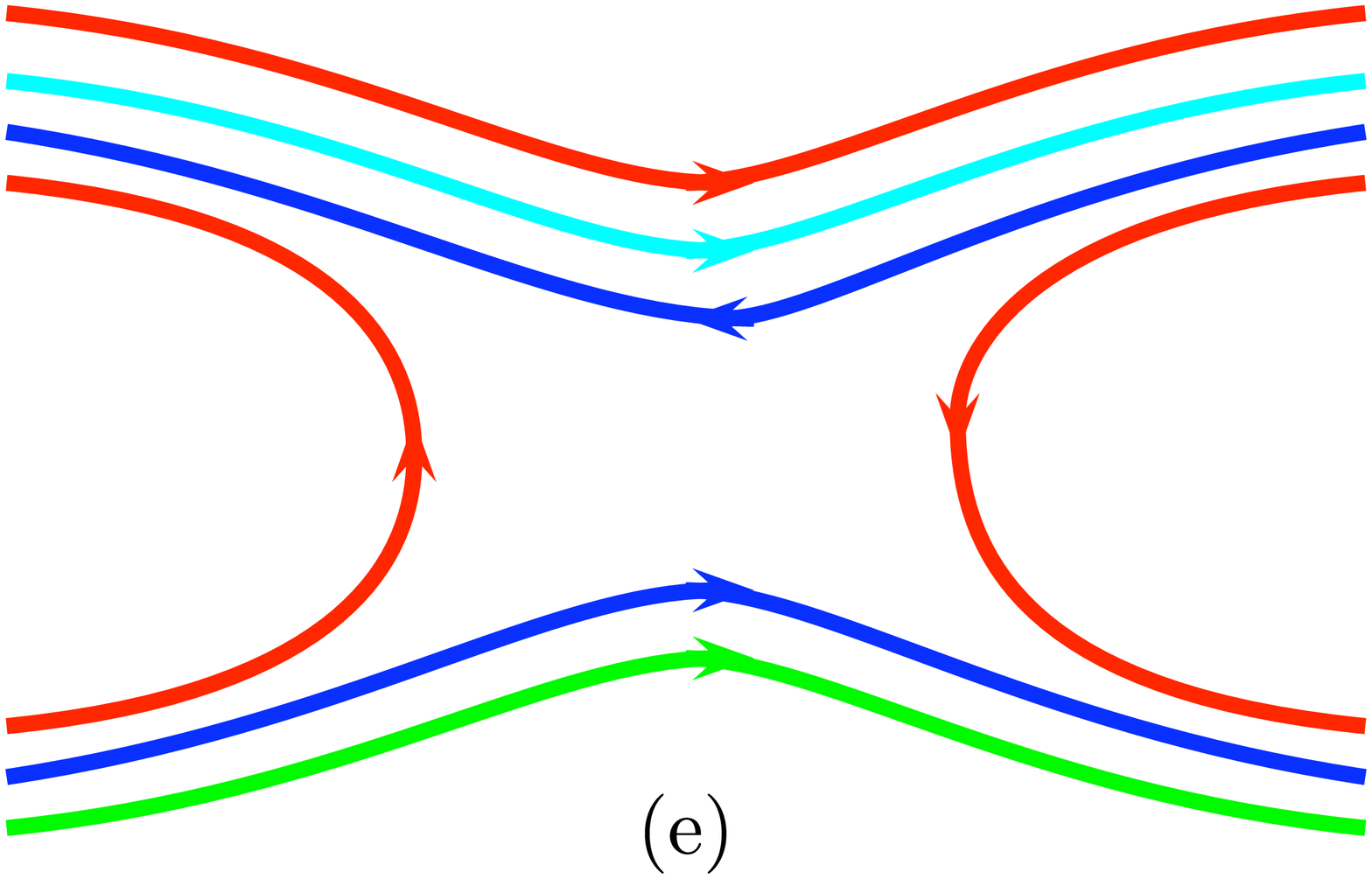}
\end{center}
\caption{\label{fig:DPRoy} $s$- and $t$-channel content for some hadronic reactions: 
(a) meson--meson, (b) meson--baryon, (c) and (d) baryon--antibaryon, and (e)  baryonium--baryon scattering.}
\end{figure}
We refer here to the  review by Roy \cite{Roy:2003hk}. The conventional approach to the 
hadron--hadron interaction was based on the Yukawa mechanism at low energy,  and Regge poles at higher energy, i.e., the exchange of particles in the $t$-channel.
The hadronic reactions without actual particles to be exchanged are thus expected to be suppressed. This is the case, for instance, of backward  $\pi^+\pi^-$ elastic scattering, seen as a forward $\pi^+\pi^-\to \pi^-\pi^+$ reaction, which involves charge $Q=2$ exchange.

On the other hand, hadronic reactions can be described by their $s$-channel content. The interaction is stronger in channels with resonances than without resonances. The consistency between these two pictures was named \emph{duality}.  The reactions with large cross sections must have actual particles exchanged in the $t$-channel and  resonances in the $s$-channel. This concept was initiated before the quark model, but became more convincing when duality diagrams were drawn in term of quark lines, as illustrated in Fig.~\ref{fig:DPRoy}.

When considering the case of nucleon--antinucleon scattering, to which ordinary mesons contribute in the $t$-channel, Rosner pointed out the need for new type of mesons, preferentially coupled to baryon--antibaryon, as their partners in the $s$-channel, see 
Fig.~\ref{fig:DPRoy}.  Later on, dynamical models were invented to tentatively describe such ``baryonium'' states, as seen below. Note that if you now enforces an ordinary meson in the $s$-channel, you need a baryonium in the $t$-channel to satisfy the duality requirements.

As pointed out in \cite{Roy:2003hk}, duality does not escape the Pan\-dora-box syndrome that  is encountered in most approaches to multiquarks. Once baryonium is accepted as a new hadron, the pentaquark automatically emerges. Consider, indeed, baryonium--baryon scattering, with  ordinary mesons  
(quark--antiquark) exchanged in the $t$-channel. Then the partner $s$-channel consists of \emph{pentaquark}, as seen in Fig.~\ref{fig:DPRoy}. Even more complicated structures appear if, in turn, a pentaquark enters a scattering with another hadron.
\subsection{Baryonium} 
In the 70s, the baryonium states anticipated by duality arguments were tentatively seen as peaks in the antiproton cross section $(M> 2\,m_p)$, or peaks in the annihilation at rest $\bar{p}p\to \gamma + X$ ($M_X< 2\,m_p$) \cite{Montanet:1980te}.  They were not confirmed, unfortunately, in later experiments with higher statistics and better resolution, in particular at the LEAR facility of CERN.

A somewhat simple picture of baryonium is that of a deuteron-like $N\overline{N}$ bound state or resonance, benefiting from  the attractive potential mediated  by the exchange of mesons \cite{Shapiro:1978wi}. The authors, however, never agreed on the influence of the very strong annihilation on the spectrum. 

Another interpretation was named \emph{diquonium} or quark-model baryonium, with two localised quarks and two localized antiquarks separated by an orbital barrier. This 
$[(qq)-(\bar q\bar q)]$ structure explains why the state is preferentially coupled to baryon--antibaryon, rather than to mesons \cite{Chan:1978nk,Montanet:1980te}. Such a clustering is, however,  assumed and not demonstrated from any dynamical calculation.

A further speculation envisaged the possibility of a colour sextet for the diquark, this leading to an even narrower $[(qq)-(\bar q\bar q)]$. This ``colour chemistry'' was also extended to the baryon number $B=1$ sector, with $[(\bar{q}q-(qqq)]$ states~\cite{deCrombrugghe:1978hi}.
\subsection{Scalar mesons}
Within  models available in the 70s (simple quark models, MIT bag), it was discovered that 
four quarks in $S$-wave,  experiencing a  favourable spin--spin interaction, have a mass which is comparable, or even lower, than an orbital excitation of $(q\bar{q})$ with a spin triplet $S=1$ and an orbital momentum $L=1$ coupled to $J=0$ \cite{Jaffe:1976ig,Jaffe:1976ih}.

 At about the same time, $K\overline{K}$ molecules were suggested \cite{Weinstein:1982gc,Weinstein:1990gu}, also candidates for scalar mesons near $1\;$GeV.  This was the beginning of an intense activity, with mixing between these various (non-orthogonal!) configurations, and further complications due to glueball and hybrid states entering the game.

The main question is whether counting the states in the spectrum  is compatible with pure $(q\bar{q})$ with higher Fock state corrections, or does require extra states. For instance, it is argued in \cite{Klempt:2007cp} that the present data in the meson sector do not call for hybrid, glueballs or molecules, except as corrections to a main $(q\bar q)$ component.

The scalar sector of light mesons is regularly revisited. In the framework of QCD sum rules, it is found that the states contain a large fraction of glueballs \cite{Kaminski:2009qg}.  A realistic description of the scalar mesons and meson--meson interaction requires the contribution of several channels \cite{Rupp:2009xg}.
\subsection{Pentaquarks}
Interesting reviews have been written on the pentaquark episode \cite{Tariq:2007ck,Wohl:2008st}. The story started with a nice speculation based of chiral dynamics \cite{Diakonov:1997mm}, leading Nakano and his colleagues to attempt a measurement out of the stream of fashion at that time, and to get a positive signal \cite{Nakano:2003qx}. The real surprise was the wave of positive evidences elsewhere, including experiments having already the data on tape for several years, that nobody had to curiosity to use to search
for exotic baryons. However, the pentaquark was not confirmed in high-statistics experiments using sophisticated and expensive detectors with good particle identification.

Some studies were also done in lattice QCD, but somewhat hastily, with a variety of conclusion about the existence of the pentaquark and about its plausible quantum numbers. Clearly, more time would have been required to  better distinguish between a genuine bound state or resonance and a state simply occurring from the discretisation of the continuum in the finite volume of the lattice.  The theoretical activity is much slowed down today, but not completely dead, in particular better studies are performed on multiquarks within the framework of QCD sum rules and lattice QCD, as seen in several contributions to this Conference, where references can be found.
\subsection{Hadronic molecules}\label{se:molecules}
This is a very  long history. In the 50s, the $\Delta$ was described as $\pi N$ resonance, i.e., a kind of \emph{molecule}. But this picture was not easily extended to other meson and baryon resonances, and when the quark model came, it was very welcome as a progress.

Today, the quark model is considered as too primitive, especially in the light-quark sector, and is given up by many authors, who describe most hadrons as molecules, again, but within the improved framework of effective theories.  Many interesting contributions have been presented at this Few-Body Conference. Any new resonance, especially in the hidden-charm sector, is almost immediately described as a new meson--meson molecule, and the Pandora box syndrome seemingly strikes again.

However, such a blame cannot be addressed to the \hbox{$X(3872)$}, since it was \emph{predicted } before its discovery, on the basis of the pion-exchange dynamics. See, e.g., \cite{Swanson:2006st} for references). If the $X(3872)$ were not confirmed as a hadron--hadron molecule and eventually interpreted as being mainly a $(c\bar{c})$ state,
we will feel  like a hen sitting on a duck egg, and rather surprised when it hatches.
We learned to be careful with the molecular interpretation of the hidden-charm states. When higher $\psi$ resonances were found, theorists were puzzled by the anomalies in the relative decay rates into $D\smash{\overline{D}}$, $D^*\smash{\overline{D}}+\mathrm{c.c.}$ and $D^*\smash{\overline{D}}{}^*$, and a molecular interpretation was suggested  \cite{DeRujula:1976qd} (see, also, \cite{Voloshin:1976ap}). But the branching ratios were later understood from the node structure of the decaying states \cite{LeYaouanc:1977ux,LeYaouanc:1977gm,Eichten:1979ms}.

As reported at this conference, the mass and the early-measured properties of the $X(3872)$ are very indicative of a $D\smash{\overline{D}}^*+\text{c.c.}$ structure. However, the branching ratio of $X\to \psi(2s)+\gamma$ to $X\to J/\psi(1s)+\gamma$ is large, a property that is natural for a radially excited P-state of $(c\bar{c})$. 

On the theory side, the problem is whether a satisfactory theory of nuclear forces can be built outside the nucleon--nucleon case.  Is has been pointed out \cite{Suzuki:2005ha}  that due to the $D-D^*$ mass difference, the pion propagator carries energy transfer together with momentum transfer, and thus that the resulting interaction is non local, and perhaps less effective.  In fact, the $X(3872)$ in this approach looks like a $D-\smash{\overline{D}}-\pi$ three-body system, with the pion oscillating back and forth between the two heavy mesons, and resonating with each of them. Another problem is that there is a hard core in the nucleon--nucleon interaction, preventing them to merge their constituents. No such core exists for $D\smash{\overline{D}}^*$ and  other meson--meson systems. Hence the short-range dynamics always plays a role. Of course, at very low binding energy, the meson--meson wave function is very extended, but the strength of the attraction is not given solely by pion exchange. the exchange of heavier mesons ($\omega, \ldots$) has been introduced (see, e.g., \cite{Swanson:2006st}), and this can be seen as a parametrisation of the direct interaction between the quarks of each flavoured meson. 
\subsection{Mixing dynamics}
This is a current trend in the physics of hadrons: if a first model describes satisfactorily some properties of a hadron, but fails for others that are accounted for by a second model, a tantalising improvement of the picture consists of writing the wave function as
\begin{equation}\label{eq:mixing}
\psi=\cos\vartheta\,\psi_1+\sin\vartheta\,\psi_2~,
\end{equation}
and this can be extended as to include more components. This game has been played endlessly  for scalar sates involving radial excitations, hidden strangeness, glueballs and hybrids. In the case of the $X(3872)$, $\psi_1$ could be a $D\smash{\overline{D}}^*+ \mathrm{c.c.}$ molecule, and $\psi_2$ a charmonium $c\bar{c}(2P)$ with the same quantum numbers and a radial excitation.  However, the dynamics of coupled channels dictates that if $\cos\vartheta\, \psi_1$ is the leading term,  the small admixture $\sin\vartheta\,\psi_2$ is \emph{not} very much governed by the diagonal interaction in the second channel. Instead it is given by a folding of the leading component and the transition operator. In particular, if the latter is nodeless and the former smooth, a node hardly appears in $\psi_2$.

A toy model illustrating this property is an S-wave analogue of the $X(3872)$ system, with a $(D\overline{D})$ (the difference between $D$ and $D^*$ is ignored) channel weakly bound by a Yukawa potential, and a $(c\bar c)$ channel with a standard Coulomb-plus-linear potential, such that before mixing, the threshold is at $3.8\;\mathrm{GeV}$, $(D\overline{D})$ at $3.77\;\mathrm{GeV}$ and $\psi(2S)$ at  $3.57\;\mathrm{GeV}$. The mixing is crudely mimicked by a local interaction with a Gaussian shape; The mass of the  $(D\overline{D})$ is slightly shifted (more details will be given elsewhere), and the wave function acquires a $(c\bar{c})$ component, displayed in Fig.~\ref{fig:coupling}, which is \emph{nodeless}. 

\begin{figure}
\begin{center}
\includegraphics[width=.40\columnwidth]{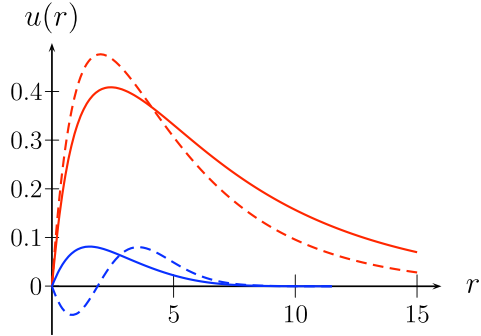}
\end{center}
\caption{\label{fig:mixing} Mixing of $(D\bar{D})$ and $(c\bar{c})$ in a simple local model. 
The solid lines represent the reduced radial wave function for the main  $(D\bar{D})$ and the small $(c\bar{c})$ components, as obtained from an actual coupled-channel calculation.  The dashed lines correspond to the naive mixing scheme of Eq.~(\ref{eq:mixing}), with neighbouring unperturbed states,  and  $\theta$ adjusted to reproduce the same normalisation for $(c\bar{c})$. Note that the coupled-channel calculation does not produce any node in the $(c\bar{c})$ sector. Units are GeV$^{1/2}$ for $u(r)$ and GeV$^{-1}$ for~$r$.}
\end{figure}

\begin{figure}
\begin{center}
\includegraphics[width=.40\columnwidth]{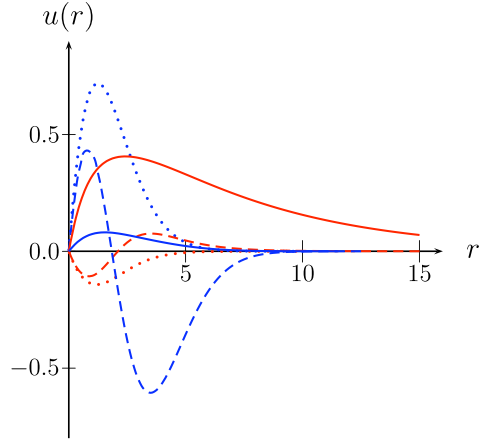}
\end{center}
\caption{\label{fig:coupled} Reduced radial wave functions for the $(c\bar{c})$ (blue) and $(D\bar{D})$ (red) components in the above coupled-channel model, for the ground-state (dotted lines), first (dashed lines) and second (solid lines) excitations.}
\end{figure}

The current belief is that the eigenstates mix unchanged, dominantly by affinity of neighbouring unperturbed energies, once a gate is open between the two channels. This is inspired by the denominator in the first order correction to a wave function within perturbation theory, namely, in a obvious notation,
\begin{equation}\label{eq:mixing}
\phi_i=\phi_i^{(0)}+ \sum_{j\neq i}\frac {\langle \phi_i^{(0)} |V| \phi_i^{(0)}\rangle}{E_i^{(0)}-E_j^{(0)}}\, \phi_i^{(0)} + \cdots
\end{equation}
However, the node structure of the $\phi_i^{(0)}$ sometimes gives more drastic constraints. Another example is $S-D$ mixing in ordinary charmonium dynamics. It is almost ever considered that $\psi(1D)$ preferentially mixes with $\psi(2S)$ which is very close. But when one actually computes the small $S$-wave admixture in $\psi(1D)$ using an explicit tensor force with suitable regularisation, one finds a nodeless radial function for the admixed  S-component in the $\psi(3770)$. Note that S--D mixing can be calculated analytically for the muonium ($\mu^+e^-$) and the approximations can be tested there.  In the above model, the ground-state has two nodeless components, with the $(c\bar{c})$ one dominating. The second state has two nodes, one in the main $(c\bar{c})$ channel, and another in the small $(D\bar{D})$  admixture, as seen in Fig.~\ref{fig:coupled}. The third one has no nodes. Many variants are possible, if the parameters are modified.

Back to $X(3872)$, the $\psi(2S)$ to $\psi(1S) $ ratio of radiative decays has to be studied within a \emph{consistent} picture involving probably several Fock-space components, $c\bar{c}$, $c\bar{c}q\bar{q}$, etc.  This requires a modelling of the transition operator creating of annihilating a pair of light quarks, and a few calculations.

A consistently-managed mixing scheme can give interesting results. An example is the celebrated Cornell picture of $(c\bar{c})$, supplemented by coupling to real of virtual $\smash{D^{(*)}}\smash{\overline{D}{}^{(*)}}$ (including $D_s$) decay channels \cite{Eichten:1979ms}. It was possible, using this model, to predict before the discovery of the $\eta_c(2s)$ a substantial reduction of the $\psi(2s)-\eta_c(2s)$ splitting \cite{Martin:1982nw}, as compared to simple potential models. The same effect probably explains why the $\Upsilon-\eta_b$ splitting is slightly larger than estimated by most authors: the reduction due  the meson--meson channels is weaker here than for $J/\psi-\eta_c$. Other interesting treatments of the configuration mixing can be found, e.g., in \cite{Gamermann:2009uq,Bali:2009er,Matheus:2009vq}.
\subsection{Diquark clustering}
The concept of diquark is very useful in several branches of particle physics, for instance to analyse the baryon-to-meson ratio in multiparticle production. 
It has been introduced rather early in spectroscopy, see, e.g., Ref.~\cite{Anselmino:1992vg} for a review.  However,  some of the pioneering works are sometimes ignored in recent rediscoveries of the diquarks.
%
%
A first problem to which diquarks provide a sufficient solution, is why Regge trajectories (squared mass $M^2$ vs.\ spin $J$) are linear with the \emph{same} slope for baryons as for mesons. For two-body systems, the linear character is reproduced in many pictures, e.g., $H_2=\sum(m^2+\vec{p}_i^2)^{1/2}+ \lambda r_{12}$.
Thus the equality of slopes comes automatically if the quark--diquark baryons, $[q-(qq)]$,  are the partners of the quark--antiquark mesons, $[q-\bar q]$.

However, it is \emph{not necessary} to introduce diquarks by hand to obtain the equality of slopes.  In the symmetric quark model, the baryon analogue of $H_2$ reads $H_3=\sum(m^2+\vec{p}_i^2)^{1/2}+ V_3$, where $V_3$ is $\lambda\,\sum r_{ij}/2$ or  the $Y$-shape potential (discussed in Sec.~\ref{se:Steiner}). The Hamiltonian $H_3$, for the large angular momentum,  gives linear Regge trajectories for baryons, with the same slope as mesons. There is a dynamical clustering, or, say, a spontaneous breaking of symmetry when $J$ increases \cite{Martin:1985hw}.

\begin{figure}[!htbc]
\begin{center}
\includegraphics[width=.50\columnwidth]{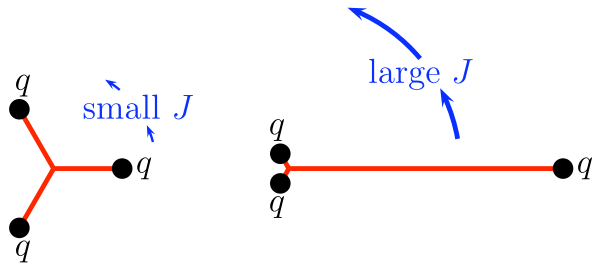}
\end{center}
\caption{\label{fig:Regge} Schematic picture of a rotating baryon at low angular momentum $J$ and higher $J$.}
\end{figure}

The best argument in favour of the quark--diquark model  perhaps lies in the problem of missing resonances. In a recent review on baryons\cite{Klempt:2009pi}, very few states are tentatively interpreted with both degrees of freedom (i.e., Jacobi variables $\vec x$ and $\vec y$) excited. On the other hand, many states predicted in the three--quark model are not observed, and this was the subject of contributions to this Conference.  The most striking state with double excitation in the three-quark model is 
\begin{equation}\label{eq:1+}
\Psi=\vec{x}\times\vec{y}\,\exp[-a\,(\vec{x}^2+\vec{y}^2)/2]~,
\end{equation}
(here in the case of harmonic interaction, but a state with the same symmetry  exists for other confining potentials). For experts, it is named the $[20,1^+]$ multiplet. Its fully antisymmetric orbital function, with $N=2$ degrees of excitation, is associated with an antisymmetric spin--isospin wave function, and an antisymmetric colour  wave function. It is absent in the quark--diquark picture. Experimentally, it is not (yet?) seen.

If the diquark model is taken seriously, it can be extrapolated outside the framework of baryon spectroscopy.  
We have seen that the late baryonium states have been described as a diquark and an antidiquark. 
More recently, $cq$ or $cs$ diquarks and the associated antidiquarks were introduced to describe the $X$, $Y$ or $Z$ states seen in the hidden-charm spectrum. See, e.g., \cite{Drenska:2009cd}.  The light pentaquark was also described as $[(qq)-(qq)-\bar{q}]$ \cite{Jaffe:2003sg}.

Obviously, the Pandora box syndrome becomes threatening here too.  In particular, three diquarks can well build a dibaryon. It should be checked whether in models describing the $Y$ mesons as $(cs)-(\bar c \bar s)$, the configuration $(cs)^3$ is not below the threshold for $(sss)+ (ccc)$, since it would be embarrassing to pay the price of a very exotic dibaryon to explain the $Y$ meson.
Years ago, a light dibaryon (or ``demon deuteron'' ) was shown to be a consequence of the diquark model  \cite{Fredriksson:1981mh}. \footnote{There is an error in  \cite{Fredriksson:1981mh} about the quantum numbers of the orbital wave function, but this does not remove the issue of dibaryons in diquark models.}%

\section{Binding mechanisms in simple quark models}\label{se:quarks}

\subsection{Chromomagnetic binding}
In the 70s, the chromomagnetic interaction,
\begin{equation}\label{eq:DiLambda}
V_{SS}=-A \sum_{i<j} \frac{\delta^{(3)}(\vec{r}_{ij})}{m_i\,m_j}\,
\lambda_i^{(c)}.\lambda_j^{(c)}\,\vec{\sigma}_i.\vec{\sigma}_j~,
\end{equation}
inspired by the Breit--Fermi interaction in atoms, was shown to account for the observed hyperfine splittings in ordinary hadrons, including the $\Sigma-\Lambda$ mass difference. There is an analogue in the bag model. 
In \cite{Jaffe:1976yi}, Jaffe pointed out the $H(uuddss)$ in its ground state has a chromomagnetic interaction whose cumulated (attractive) strength is larger than for the lowest threshold $\Lambda(uds)+\Lambda(uds)$.  Note the \emph{non-trivial} colour--spin algebra. In a model with pure spin--spin, $V_{SS}\propto \sum \vec{\sigma}_i.\vec{\sigma}_j$, a $J=0$ tetraquark receives a cumulated strength
$\sum_{i<j}\vec{\sigma}_i.\vec{\sigma}_j=-6$, whereas \emph{each} scalar meson of its threshold experiences $\sum_{i<j}\vec{\sigma}_i.\vec{\sigma}_j =-3$, and thus there is no obvious excess of attraction.  At best a small gain from a favourable  spatial distortion.
Remember also the positronium molecule, governed by $V=\sum_{i<j} g_{ij}/r_{ij}$: its binding is not obvious because the cumulated strength is 
$\sum g_{ij}=-2$ for the molecule, and $\sum g_{ij}=-2$ also for the threshold made of two atoms.
The miracle in (\ref{eq:DiLambda}) is that
 \begin{equation}\label{eq:CM}
 \bigl\langle \lambda_i^{(c)}.\lambda_j^{(c)}\,\vec{\sigma}_i.\vec{\sigma}_j\bigr\rangle_H
 =2\, \bigl\langle \ldots \bigr\rangle_{\rm threshold}
 \end{equation}

If the quark masses $m_i$ are identical  ($\mathrm{SU(3)_f}$ symmetry) and if the short-range correlation factors $\langle \delta^{(3)}(\vec{r}_{ij}) \rangle$ are assumed to be the same in multiquarks as in ordinary baryons, then the $H$ is predicted about $150\;$MeV below  the threshold. 
The $H$ was searched for in more than 15 experiments, with negative results, including double-$\Lambda$ hypernuclei.

In 1987, Lipkin, and independently Gignoux et al.\  \cite{Lipkin:1987sk,Gignoux:1987cn,Lipkin:1998pb} pointed out that the same mechanism would bind an exotic pentaquark (the word was invented in this context)
 \begin{equation}\label{eq:P}
P(\overline{Q}qqqq)< (\overline{Q}q)+(qqq)~,
\end{equation}
where $Q=c,\, b$ and $(qqqq)$ is a combination of $u$, $d$, $s$, in a SU(3)$_{\rm f}$ flavour triplet. The same binding of $-150\;$MeV is found if $Q$ becomes infinitely heavy. 
A search for the $P$ r in an experiment at Fermilab turned out not conclusive \cite{Aitala:1997ja,Aitala:1999ij}. Silvestre-Brac and Leandri listed other candidates \cite{SilvestreBrac:1992yg}.

However, further studies  of the chromomagnetic model \cite{Rosner:1985yh,Karl:1987cg,Fleck:1989ff,Sakai:1999qm} indicated that the corrections tend to moderate and even to cancel the binding effect due to the group-theoretical result (\ref{eq:CM}). In particular:
\begin{itemize}
\item There is more kinetic energy in a dibaryon than in two baryons (5 Jacobi variables vs. 4),
\item SU(3)$_{\rm f}$ breaking is not favourable. The $\Lambda$ resists much better than $H$.
\item The matrix elements $\langle \delta^{(3)}(\vec{r}_{ij}) \rangle$ appreciably smaller in multiquarks than in ordinary hadrons.
\end{itemize}

Nevertheless, the simple chromomagnetic Hamiltonian (\ref{eq:DiLambda}) might still reveal some surprises. For instance, if diagonalised with realistic values for the flavour dependent strength factors $\langle \delta (\vec{r}_{ij}\rangle/(m_im_j)$, it provides the $(c\bar{c} q\bar{q})$ lowest configuration with $J^{PC}=1^{++}$ the main features required to describe the $X(3872)$ \cite{Hogaasen:2005jv,Buccella:2006fn}.

\subsection{Chromoelectric binding}
We adopt here the language of potential models but we believe that the results are much more general. The main advantage of potential models is the possibility of switching on or off some contributions to single out the most effective one for binding. 

A remarkable property of the spin-independent interaction among quarks is \emph{flavour independence}, which induces interesting \emph{symmetry breaking} effects.

Remember that symmetry breaking tends to lower the ground-state energy. The simplest example is $h=p^2+x^2+\lambda\,x$ in one-dimensional quantum mechanics, with a  ground state energy $e(\lambda)=1-\lambda^2/2$ which  is always below $e(0)$. But this is very general. If
\begin{equation}\label{eq:sb1}
H(\lambda)=H_\text{even}+\lambda\,H_\text{odd}~,
\end{equation}
 the variational principle applied to $H(0)$ with  the even ground state $\Psi(0)$ of $H_\text{even}$ as trial wave function gives
\begin{equation}\label{eq:sb2}
E(\lambda)\le \langle \Psi(0) | H(\lambda) | \Psi(0)\rangle = E(0)~.
\end{equation}
One can apply this result to few-body systems with a variety of symmetries for which the labels ``odd'' and ``even'' make sense, in particular \emph{particle identity} and \emph{charge conjugation}. But stability is a competition between a configuration with collective binding and another configuration where the system is split into separate decay products. The threshold also benefits from the symmetry breaking, and very often more! In this case, stability deteriorates, though the energy of the compound configuration decreases.

For instance, consider  the barely bound $(e^+,e^+,e^-,e^-)$ molecule, or any rescaled version with the electron mass replaced by another mass $\mu$, and move to
configurations involving two different masses. Then it is observed, and proved, that:
\begin{itemize}
\item for $(M^+,m^+,M^-,m^-)$:  binding deteriorates and is lost for $M/m\gtrsim 2.2$ \cite{PhysRevA.57.4956},
\item for $(M^+,M^+,m^-,m^-)$: the binding is improved.
\end{itemize}
But the Coulomb character matters little. What is important, is that the potential does not change when the masses are varied.  Hence, a similar behaviour is observed for any four-body problem with \emph{flavour independence}. The splitting (the potential $V$ is assumed to be symmetric under charge conjugation, and independent of the masses)
\begin{eqnarray}\label{eq:sb3}
&&H(M,M,\bar m,\bar m)=\frac{\vec p_1^2}{2 M}+\frac{\vec p_2^2}{2 M}+
\frac{\vec p_3^2}{2 m}+\frac{\vec p_4^2}{2 m}+ V~,\\
&&\ =H(\mu,\mu,\bar\mu,\bar\mu)+\left[\frac{1}{4M}-\frac{1}{4m}\right]
(\vec p_1^2-\vec p_2^2+\vec p_3^2-\vec p_4^2)~,\nonumber
\end{eqnarray} 
implies for the ground state
\begin{equation}\label{eq:sb4}
E(M,M,\bar m,\bar m)\le E(\mu,\mu,\bar\mu,\bar\mu)~, \quad 2\,\mu^{-1}=M^{-1}+m^{-1}~,
\end{equation}
but, meanwhile, the threshold energy remains constant, as $(M,\bar m)$ and $(\mu,\bar\mu)$ have the same reduced mass. So the stability is improved.

Explicit quark-model calculations have been carried out to illustrate how this favourable symmetry breaking works with flavour-independent potentials. 
The corresponding four-body problem is rather delicate, as most other four-body problems. Remember that when Wheeler proposed in 1945 (the paper paper was published somewhat later \cite{Wheeler46a}) that the positronium molecule might be stable, a first numerical investigation by Ore \cite{PhysRev.70.90} concluded that the system is likely unstable, but the following year, Hylleraas and the same Ore published a beautiful analytic proof of the stability \cite{PhysRev.71.493}.

In current quark models, the main conclusion is that
binding a doubly-flavoured tetraquark
requires a large mass ratio, usually $(bb\bar{q}\bar{q})$ or $(bc\bar{q}\bar{q})$.  However, a more sophisticated calculation by Janc and Rosina \cite{Janc:2004qn} found $(cc\bar{q}\bar{q})$ barely bound. See, e.g., \cite{Vijande:2009kj} for a detailed survey of the situation.


\section{Steiner-tree  model of confinement}\label{se:Steiner}
It should be acknowledged, however, that these early constituent-model calculations suffer from a basic ambiguity: how to extrapolate from mesons towards multiquarks. The usual recipe is 
\begin{equation}\label{eq:col-add}
V=-\frac{3}{16}\,\sum_{i<j}\tilde\lambda^{(c)}_i.\tilde\lambda^{(c)}_j\,v(r_{ij})~,
\end{equation}
which is presumably justified for the short-range part, but not for the long-range part, except for very peculiar spatial configurations. In (\ref{eq:col-add}), the normalisation is such that $v(r)$ is the quark--antiquark interaction in ordinary mesons.
Strictly speaking, (\ref{eq:col-add}) holds for a \emph{pairwise} interaction with colour-octet exchange. Clearly colour-singlet exchange cannot contribute to confinement, otherwise everything would be confined together, but colour-singlet  exchange can contribute to short-range terms. Moreover, there are very likely three-body and multi-body forces in baryons and multiquarks. 

\begin{figure}[hbtc]
\centerline{
\includegraphics[width=.20\columnwidth]{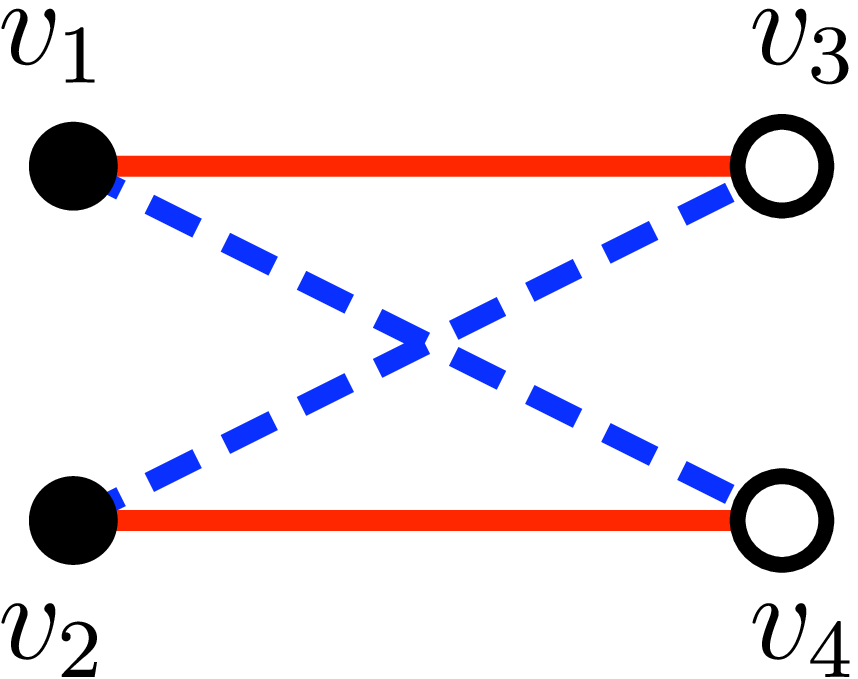}\hfil
\includegraphics[width=.20\columnwidth]{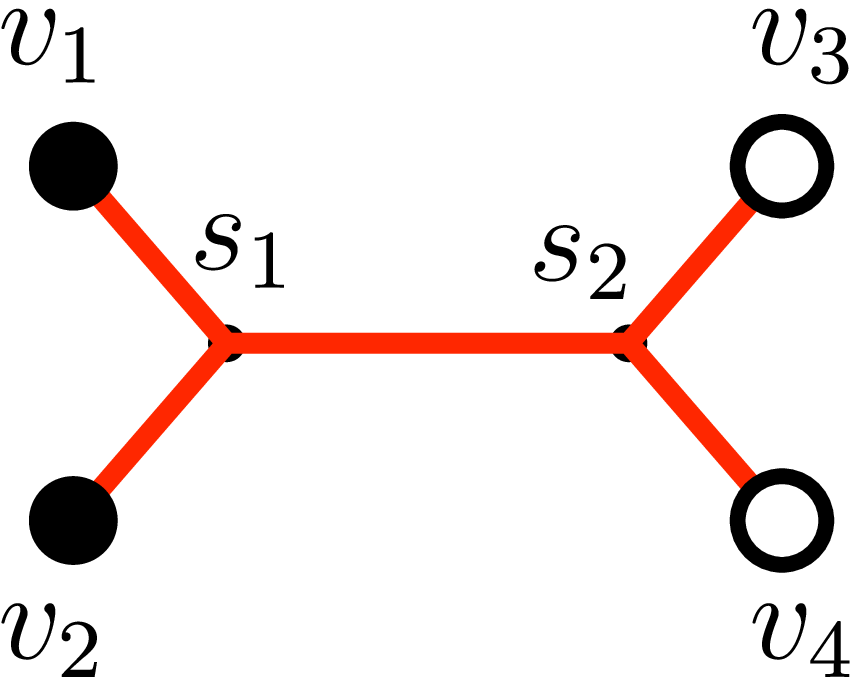}
}
\caption{\label{fig:bar-tetra} Generalisation of the linear potential of mesons to  tetraquarks:  the minimum is taken of the flip--flop (left) and Steiner tree (right) configurations.}
\end{figure}

In the case of baryons, it was suggested very early \cite{Artru:1974zn,Dosch:1975gf} that the potential generalising the linear confinement of mesons is 
\begin{equation}\label{eq:VY}
V_Y=\sigma\,\min_J\sum_{i=1}^3 r_{iJ}~.
\end{equation}
This was often rediscovered in the context of models (adiabatic bag, flux tubes), or in studies dealing with the strong-coupling regime of QCD \cite{Takahashi:2000te}. 
\begin{figure}
\begin{center}
\includegraphics[width=.42\columnwidth]{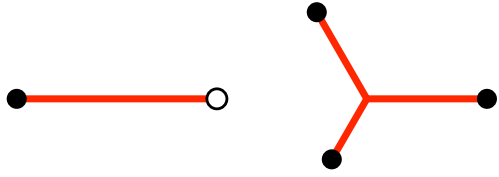}
\end{center}
\caption{\label{fig:Y} Confinement of mesons and baryons. The minimum over the quark permutations gives the flip--flop potential.}
\end{figure}
Estimating the baryon energies and properties with the potential (\ref{eq:VY}) is a very interesting 3-body problem. However, the result differ little from these obtained using  the colour-additive rule, which for baryons reduces to the ``1/2" rule, namely
\begin{equation}\label{eq:Vhalf}
V_3=\frac{\sigma}{2}\left( r_{12}+r_{23} +r_{31}\right)~.
\end{equation}
This $Y$-shape interaction has been generalised to tetraquarks. At first, this looks as an astute guess, but  it was later endorsed by detailed lattice QCD \cite{Okiharu:2004ve}, including the interplay between flip--flop and connected Steiner tree. See, also, \cite{Andreev:2008tv} for a study within AdS/QCD. The confining potential reads
\begin{eqnarray}\label{tetra-pot}
U&=&\min \left\{ d_{13}+d_{24},d_{14}+d_{23},V_4 \right\}~,\\
\qquad
V_4&=&\min_{s_1,s_2}\left(\, \|v_1s_1\| + \| v_2s_1\|+ \| s_1s_2\|+\| s_2v_3\|+\| s_2v_4\|\,\right )~,\nonumber
\end{eqnarray}
corresponding schematically to the flux tubes in Fig.~\ref{fig:bar-tetra}

A first study of the tetraquark spectrum with this potential concluded to the
 ``absence of exotics'' \cite{Carlson:1991zt}, but a re-analysis by Vijande et al.\ \cite{Vijande:2007ix} with a better wave function, indicated that this potential, if alone, and free of constraints due to the Pauli principle, gives stability for the equal-mass case $(qq\bar{q}\bar{q})$, and improved stability for the flavour asymmetric $(QQ\bar{q}\bar{q})$. It remains to analyse how this stability survives antisymmetrisation, short-range terms in the potential, relativistic effects and spin-dependent corrections. This is however, very encouraging.

\begin{figure}
\centerline{\includegraphics[width=.28\columnwidth]{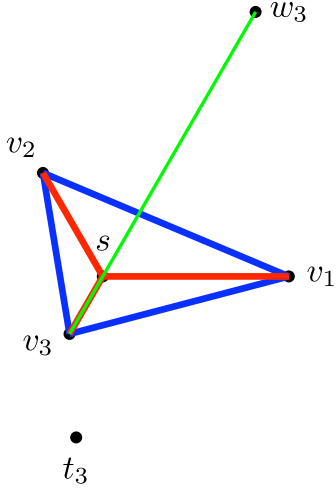}}
\caption{\label{fig:Melzak3} The continuous minimisation over the location of $s$ is replaced by the discrete maximum $\max\left\{\|sw_3\|,\, \|st_3\|\right\}$, where $w_3$ and $t_3$ make an equilateral triangle with the quarks $v_1$ and $v_2$.
Here, $V_3=\| sv_1 \| + \| sv_2 \| + \| sv_3 \| = \| v_3 w_3 \|$.}
\end{figure}

Two developments came. First, a better understanding (at least by physicists) of the Steiner trees, with efficient algorithms \cite{Bicudo:2008yr} to compute them, and novel inequalities \cite{Ay:2009zp}. For instance, in a Steiner tree, a set of point can be replaced by its associated torro\" idal domain, 
For instance, in the baryon $Y$-problem, the continuous minimisation over the location of the Steiner point $s$ can be replaced by a \emph{discrete maximisation}: the desired $Y$-shape length, $\| sv_1 \| + \| sv_2 \| + \| sv_3 \|$ in Fig.~\ref{fig:Melzak3} is the largest of the two distances $\|sw_3\|$ and  $\|st_3\|$, where the points $w_3$ and $t_3$, named the Melznak points or torro\"idal domain of $\{v_1,v_2\}$, form an equilateral triangle with $v_1$ and $v_2$. 

\begin{figure}
\centerline{\includegraphics[width=.40\columnwidth]{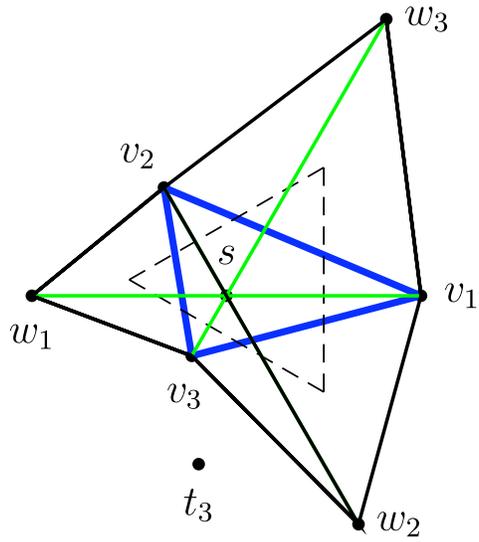}}
\caption{\label{fig:Napoleon} Baryon potential and Napoleon's theorem.}
\end{figure}

This property, and the interesting \emph{symmetry restoration} (even for an asymmetric triangle, the flux tubes from the junction to the quarks make $120^\circ$ angles in Fig.~\ref{fig:Y}) are related to the theorem by Napoleon (see Fig.~\ref{fig:Napoleon}): if one starts from an asymmetric triangle $v_1v_2v_3$ and builds  an external equilateral triangle such as $v_1v_2w_3$ along each side, the centres of these auxiliary triangles form an equilateral triangle.

\begin{figure}
\centerline{\includegraphics[width=.60\columnwidth]{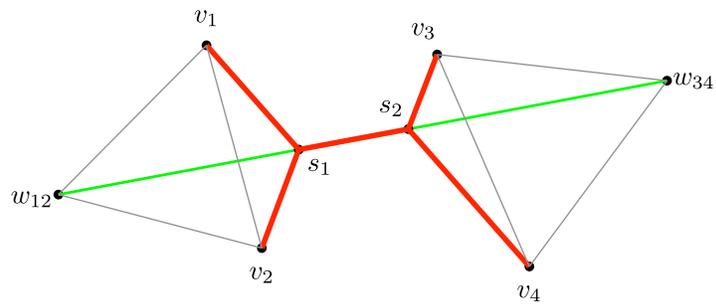}}
\caption{\label{fig:tetrapl} The length of the Steiner tree is the distance between the vertices making an external triangle in the quark and in the antiquark sector.}
\end{figure}

The analogue for a planar  tetraquark is 
\begin{equation}\label{eq:tetrapl}
V_S=\| s_1 v_1 \| + \| s_1 v_2 \|+ \| s_2 v_3 \|+ \| s_2 v_4 \|= \| w_{12} w_{34} \|~,
\end{equation}
and is illustrated in Fig.~\ref{fig:tetrapl}: the length of Steiner tree linking the quarks $v_1$ and $v_2$ to the antiquarks $v_3$ and $v_4$ \textsl{via} the junctions $s_1$ and $s_2$ is he distance between the two Melznak points $w_{12}$ and $w_{34}$.

\begin{figure}
\centerline{\includegraphics[width=.35\columnwidth]{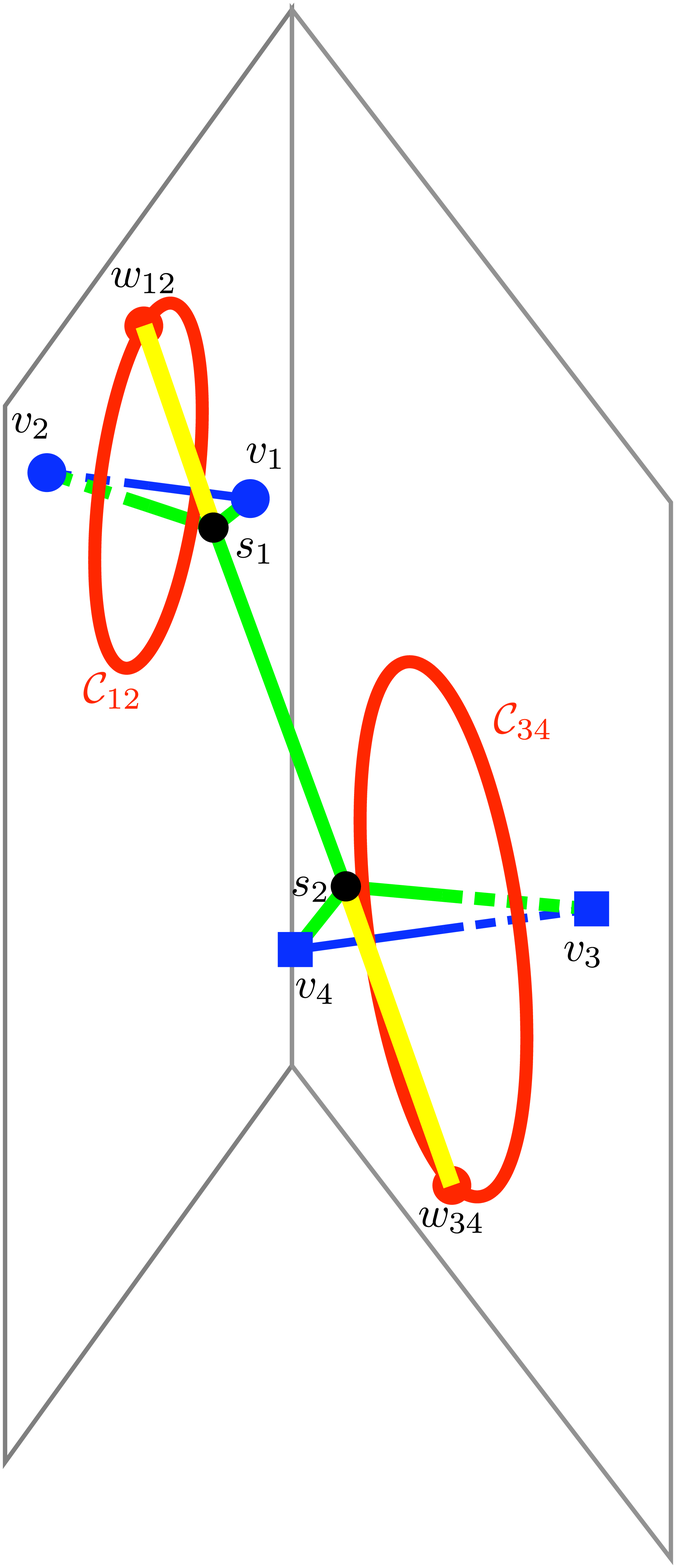}}
\caption{\label{fig:tetrasp} The length of the Steiner tree is the maximal distance between the quark and antiquark Melznak circles.}
\end{figure}

In space, one gets the problem pictured in Fig.~\ref{fig:tetrasp}: the auxiliary points $w_{12}$ and $w_{34}$ are located on a circle of axis $v_1v_2$ (resp.\ $v_3v_4$), and estimating the minimal Steiner tree can be shown to be equivalent to finding the \textsl{maximal} distance between the two circles \cite{Ay:2009zp}. This is a standard problem in computer-assisted geometry, as applied, e.g., in the cartoon industry.

Then you can derive inequalities on the Hamiltonian, and recover rigorously in  some limiting cases the stability established by numerical methods.

The disappointing observation, however, is that the dynamics is dominated by the simple \emph{flip--flop} term, and the most interesting connected Steiner tree plays a relatively minor role.

\begin{figure}[!htbc]
\begin{center}
\includegraphics[width=.40\columnwidth]{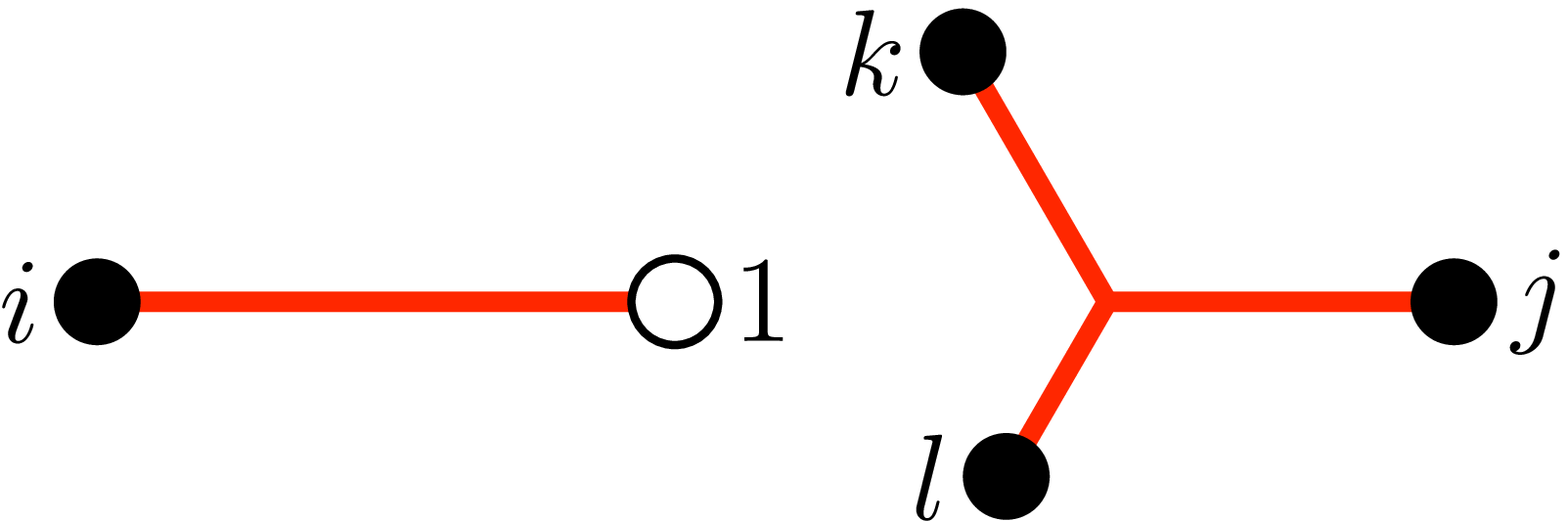}\\[10pt]
\includegraphics[width=.32\columnwidth]{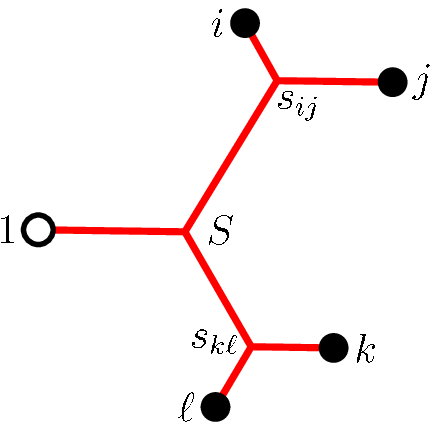}
\end{center}
\caption{\label{fig:penta} Pentaquark confining potential: flip--flop (up) and connected (down).}
\end{figure}

The exercise can be repeated for the pentaquark, using the linear model with minimal cumulated length, 
\begin{eqnarray}\label{eq:Vff}
V_P&=&\min(V_\text{ff},V_\text{St})~,\nonumber\\
V_\text{ff}&=&\min_i\left[r_{1i}+ V_Y(\vec{r}_j,\vec{r}_k,\vec{r}_\ell)\right]~,\\
V_\text{St}&=& \text{connected\ Steiner\ tree}~,\nonumber
\end{eqnarray}
i.e., the configurations in Fig.~\ref{fig:penta}.

A simple variational calculation gives stability at least for $(\bar{q}qqqq)$, and $(\overline{Q}qqqq)$ and $(\bar{q}qqqQ)$ where $m(Q)\gg m(q)$ \cite{Richard:2009rp}. The other mass configurations remain to be studied. 

This proliferation of stable states in the minimal-length model becomes embarrassing. Very likely, the dibaryon will also be found stable. If one believes into this confinement, this means that the role of the neglected effects should be investigated with care, in particular:
\begin{itemize}
\item Coulomb forces, perturbative and non perturbative,
\item antisymmetrisation.
\end{itemize}
If the constraints of antisymmetrization turn out the main obstacle to multiquark stability, then exotic hadrons have to be searched in configurations with quarks of different flavours. 
\section{Conclusions}\label{se:out}
The problem of multiquark binding is now addressed very seriously with QCD sum rules, Lattice QCD and even AdS/\-QCD. These ambitious but delicate approaches have first confirmed some results that were previously obtained empirically, such as the Steiner-tree structure of the linear term of the confining interaction. 

Constituent quark models remain, however, in the forefront of investigations, to detect the most interesting configurations. With a proper treatment of relativistic effects, and adequate dynamics, quark-model calculations can give a very good account of the spectrum and main static properties, as illustrated, e.g., by the Bonn group for mesons and baryons \cite{Metsch:2008zz}.  The case of multiquark is of course much more difficult, with the mixing of confined channels and hadron--hadron components in the wave function. 

On the experimental side, it is hope that the future collider experiments will devote a reasonable amount of time to search for exotics with heavy flavour. As shown by $B$ factories, there is a very good potential of discoveries in hadron physics within experiments primarily designed for studying other aspects of particle physics.

\subsubsection*{Acknowledgments} It is a pleasure to thank the organisers of this beautiful Conference for giving us the opportunity of many stimulating discussions.
%
%

\end{document}